\documentclass[useAMS,usenatbib]{mn2e}
\usepackage{graphicx}

\newcommand{\mch}{$\mathrm{M_{Ch}}$}
\newcommand{\mche}{$\mathrm{M_{Ch}}\ $}
\newcommand{\msun}{$\mathrm{M_\odot}$}
\newcommand{\msune}{$\mathrm{M_\odot}\ $}
\newcommand{\medd}{$\mathrm{\dot{M}_{Edd}}$}
\newcommand{\msunyrm}{$\mathrm{M_\odot yr^{-1}}$}
\newcommand{\mdot}{$\mathrm{\dot{M}}$}
\newcommand{\mdote}{$\mathrm{\dot{M}}\ $}

\title[Pre-Explosive properties of SNe Ia]{Pre-Explosive Observational Properties of Type Ia Supernovae}
\author[A. Tornamb\'e and L. Piersanti]
{A. Tornamb\'e$^{1}$\thanks{E-mail:
tornambe@oa-teramo.inaf.it (AT); piersanti@oa-teramo.inaf.it (LP)}, L. Piersanti$^{2}$\\
$^{1}$INAF - Osservatorio Astronomico di Roma, via di Frascati, 33, 00040, Monteporzio Catone - ITALY\\
$^{2}$INAF - Osservatorio Astronomico di Teramo, via Mentore Maggini, snc, 64100, Teramo - ITALY }
\begin{document}

\date{Accepted ... Received ...; in original form ...}

\pagerange{\pageref{firstpage}--\pageref{lastpage}} \pubyear{2012}

\maketitle

\label{firstpage}

\begin{abstract}
The evolutionary path of rotating CO White Dwarfs directly accreting CO-rich matter 
is followed up to few seconds before the explosive breakout in the framework
of the Double Degenerate rotationally-driven accretion scenario. 
We compute several models with different initial 
masses and physical conditions, following all the evolutionary phases, from the 
heating process during the pre-merging phase, through the two self-regulated
accretion phases, up to the final central C-ignition and the development of an 
extended convective core. 

We find that the evolutionary properties (both structural and observational) depend 
only on the actual mass of the accreting White Dwarf and not on the previous history. 
We determine the expected frequency and amplitude of the gravitational wave 
emission, which occurs during the mass transfer process 
and, as a matter of fact, acts as a self-tuning mechanism of the accretion process 
itself. 
The gravitational signal related to Galactic sources can be easily detected with the 
next generation of space-born interferometers and can provide notable constraints to 
the progenitor model. 
The expected statistical distribution of pre-explosive objects in the Galaxy is 
provided also in the effective temperature-apparent bolometric magnitude diagrams 
which can be used to identify merged DD systems via UV surveys, once the contribution 
of the accretion disk is properly taken into account.

We emphasize that the thermonuclear explosion occurs owing to the decay of physical 
conditions keeping over-stable the structure above the classical Chandrasekhar limit 
and not by a steady increase of the WD mass up to this limit. This conclusion is 
independent of the evolutionary scenario for the progenitors, but it is a direct 
consequence of the stabilizing effect of rotation. 
Such an occurrence represents an epistemological change of the perspective in defining 
the ignition process for accreting WDs. Moreover, this requires a long evolutionary 
period (of at least several million years) to attain the explosion after the above 
mentioned conditions cease to keep stable the WD.
Therefore it is practically impossible to detect the trace of the exploding WD 
companion in recent pre-explosion frames of even very near SN Ia events. 
\end{abstract}


\section{Introduction}\label{s:intro}
The recent discovery of the type Ia Supernova (SN Ia) SN2011fe in the nearby galaxy 
M101 has reawakened the interest to get direct information on the progenitor 
systems of these explosive events, searching in archive images any signature 
to distinguish among various progenitor models (see, {\it e.g.}, 
\citeauthor{bloom2012} \citeyear{bloom2012}). 
In principle, once the scenario for the progenitors has been fixed, the evolutionary 
path up to the explosion can be modeled and the results compared with the 
observational data obtained before explosion. 
{\bf 
In recent years, this approach has been adopted by many authors, even if no definite 
conclusions on the nature of SNe Ia progenitors has been obtained \cite[see][]
{mao2008,voss2008,niels2012,li2011b}. 
}

Both the two most common scenarios for type Ia SNe progenitors, the Double Degenerate 
(DD) and the Single Degenerate (SD) scenarios, require that rotation plays an 
important role in the evolution of accreting WDs 
{\bf
(\citeauthor{pie2003b} \citeyear{pie2003b} - hereinafter PGIT,  
\citeauthor{diste11} \citeyear{diste11}, \citeauthor{hac12} \citeyear{hac12}). 
}
This is true also for the Core Degenerate scenario by \cite{kashi2011}, 
assuming that a CO WD merges with the core of AGB star soon after the end 
of a Common Envelope phase. 
In this case an evolutionary time lasting more than $10^6$ yrs is expected from the 
halting of the accretion process up to C-ignition in highly degenerate physical 
conditions and, after an additional short time related to the simmering phase, 
to the explosion. 
This implies, as will be discussed later on, that it is not possible to find 
in archival images any hints about the typology of the donor star, but it will be possible 
to observe only the exploding star, whose properties are almost independent of the 
considered scenario for the progenitors.
As a matter of fact the only possibility to derive information about the mass 
transferring companion (that is the progenitor model) is to detect observationally 
some features characterizing the evolving system during the mass accretion 
phase and beyond, up to the explosion. 
 
In the present work we focus our attention on the Double Degenerate scenario, 
assuming that SNe Ia arise from binary systems made by two CO White Dwarfs (WDs) with 
total initial mass larger than the canonical Chandrasekhar mass limit \mche and 
with an initial orbital separation small enough 
to allow the merging of the two components via gravitational wave radiation (GWR) on 
a timescale smaller than the Hubble time \citep{iben84,web84}. According to 
\cite{benz90} and \cite{rasio95} the less massive component overfills its own Roche lobe 
and completely disrupts. The debris of this WD forms an accretion disk 
from which matter flows to the more massive survived companion \citep{tut79}.

One first shortcoming for this scenario is the conjecture that during the merging 
process carbon could be ignited near the surface of the accreting star and the 
burning transferred very rapidly towards the inner regions, thus producing as a 
penultimate outcome a O-Ne-Mg WD. The latter will eventually collapse into a neutron 
star \citep{isern83,her88}, producing an explosive event, but not a thermonuclear 
Supernova. This has been excluded by \cite{lig2009}, who find that early C-burning 
indeed occurs, but it is soon extinguished. 

The second important shortcoming is that, after the merging, the mass transfer
from the disk to the surviving WD occurs at a very high 
rate, close to the Eddington limit (\medd$\sim 10^{-5}$ \msunyrm), 
so that off center C-burning has to occur at the base of the 
accreted layers well before the WD could attain \mche \citep{saio85,saio98}. 

However, PGIT have shown that, if the effects of rotation, which naturally 
arise in merging DD systems, are taken into account, the accretion process 
becomes ``self-regulated'', producing, at the end, a degenerate central carbon 
ignition and a type Ia SN-like outburst. 
Another important result of the rotating Double Degenerate scenario is that the 
total mass of the accreting WD can increase above the canonical \mch, up to the 
corresponding mass limit for rotating degenerate objects. 
If for the accreting WD very high efficiency of angular momentum transport is assumed 
(rigid body rotation) as in PGIT, such a limit is $\sim 1.5$\msun, but it could be  
larger, up to $\sim 4$\msun, for differentially rotating objects, depending on the 
angular momentum distribution \citep{ostra68}. 
However, it has to be noted that the maximum total mass of binary CO WDs systems 
is limited in nature to $\sim 2.2 - 2.4$ \msune due to evolutionary constraints.

As first suggested by \cite{pal2009}, the accreting CO WD will naturally evolve 
to the explosion, once it has accreted over the classical \mche 
in an over-stable condition, independently of the assumed efficiency of the angular 
momentum transport in the accreting structure, either because angular momentum is 
lost from the star by viscous friction (rigid rotating body) 
or because angular momentum is redistributed through the structure via viscous shear 
(differentially rotating body).

We remark that rotation, besides acting 
as the leading parameter of the accretion process, implies also a deep change of 
perspective in our understanding the evolution up to the explosion. 
In fact we are no longer faced with a matter-accreting WD simply growing to a 
mass limit and, hence, experiencing a strong compression which triggers the explosion. 
By contrast, we face with an over-stable rotating structure still increasing in mass 
also after the mass limit has been largely exceeded, because the actual total mass 
remains smaller than the {\sl rotating} \mch. Such an occurrence may allow all the mass 
available in the binary system to be accreted onto the WD. 
Later on, owing to either the loss or the 
redistribution of angular momentum, the value of the rotating \mche decreases 
and only when it approaches the actual total mass of the WD, 
the whole structure strongly contracts, producing the central degenerate C-ignition.
Such a process reminds of what occurs in the degenerate cores of massive stars
at the onset of the collapse, when, owing to deleptonization, \mche decreases below 
the actual value of the core itself and not vice-versa.

The evolutionary scenario proposed by PGIT provides a detailed description of the 
physical properties of the accreting WD from the merging phase 
up to the Supernova explosion, thus allowing the definition of observable quantities 
which can be used to put more firm constraints on the SN progenitor 
systems. A similar approach, but employing
an even earlier evolutionary phase, has been followed recently by \cite{bad+mao012}.

{\bf 
In \S \ref{s:valida} we review the scenario for the merging of two CO WDs and 
we address the problem of the long term evolution of the merged object. 
We also review the input physics and the numerical procedures 
adopted to compute evolutionary sequence of accreting WDs according to the 
PGIT prescriptions. The assumptions and their implications 
are also discussed.}
In \S \ref{s:evolu} we discuss the surface observable properties of accreting WDs 
during the whole process, from the merging up to the Supernova explosion.
This has been done for various original masses of the accreting White Dwarf.
{\bf 
According to the obtained results we identify some key evolutionary properties 
of accreting WDs and we demonstrate that they are independent on the initial physical 
conditions of the post-merging object.
}
We define also the properties of the gravitational signal related 
to the self-regulated accretion phase. 
In \S \ref{s:synthe} we determine for the Galaxy the expected number objects experiencing the 
various accretion phases and we also define the expected signals, both in the 
electromagnetic and in the gravitational wave radiation domain, 
In our analysis we ignore all those systems with a total mass too small to provide an explosive 
outcome (``SN manque'' systems), focusing our attention only on DD systems good 
candidates as SNe Ia progenitors.
Our conclusions are summarized in \S \ref{s:conclu}.

\section{Merging WDs: an Overview}\label{s:valida}

The merging process of two White Dwarfs has been discussed by many authors 
during the last three decades, even if it has not bween attained so far 
a firm and widely accepted conclusion on the outcomes.

After the pioneering works by \cite{benz90} and \cite{rasio95}, a re-analysis 
of the merging phase of two CO WDs has been performed using SPH simulations 
by \cite{guer2004}, \cite{ross2007}, and \cite{lig2009}. The obtained 
results confirm that the less massive component completely destroys in a 
dynamical timescale, forming an hot and thick accretion disk around the surviving
companion. 
The timescale on which the merging occurs has been the subject of several dicussions
(see, e.g., \cite{dsou2006} and \cite{motl2007} \cite{dan2011}\cite{jor2012}
\cite{raskin2012}). Currently, a general consensus does exist on the fact that the 
merging takes several orbits to occur and, in any case, when the merging process 
starts, \mdote rapidly increases, driving to the formation of an hot corona 
surrounded by a thick disk around the WD companion.

At contrast, \cite{pakmor2010,pakmor2011} suggested that the DD systems made by 
two CO WDs more massive than $\sim 0.9$\msune and with mass ratio larger than 0.8 
undergo a violent merging. \cite{pakmor2012}
computed the full evolution of a 1.1+0.9 \msune initial binary system, from 
the merging phase through a prompt thermonuclear explosion. 
Such a finding has been questioned by \cite{dan2011} and \cite{raskin2012}, 
even if they show that a detonation occur for sure if the destroyed component has a 
large enough pure He buffer on the surface or it is an He WD. 

\cite{shen2012} and \cite{schwab2012} analize the long term evolution of the 
post-merging configuration and they claim that
the disk formed during the merging rapidly disperses 
its angular momentum, thus evolving into an expanded, slowly rotating, almost 
spherical envelope. In this case off-center carbon ignition will occur 
either during the merging or when the expanded envelope relaxes and contracts 
thus producing an O-Ne-Mg core which eventually 
will collapse, if the total mass of the initial binary is larger than \mch.
However, in this scenario the accretion on the central object is inhibited thus 
forcing the the disk-like configuration to dissipate its angular momentum. 
Therefore, the results by \cite{shen2012} can not be used to exclude the 
formation of a keplerian disk and the consequent accretion at \medd \ onto the 
survived CO WD.

The post-merging evolution of a CO WD plus a thick accretion disk has been 
analyzed first by \cite{nomoto1985} and \cite{saio85,saio98} assuming 
\mdot$\sim$\medd \ and
neglecting the effect of rotation. According to their results, C-burning is 
ignited off-center and it propagates inward up to the center, producing a 
O-Ne-Mg WD which eventually could collapse into a neutron star \citep{isern83,her88}.
The effect of rotation in the evolution of merged CO WDs was included for the 
first time by \cite{pie2003a}. Basing on this systematic investigation of the 
thermal response of the accreting WD on the mass and angular momentum deposition, 
PGIT proposed an evolutionary model describing the evolution of the whole 
accretion process up to the C-ignition at the center in high degenerate physical 
conditions. PGIT show that, due to the continuous mass accretion at a very 
high rate and the deposition of angular momentum, the accreting WD experiences 
a Roche instability at the equator. As explained in \cite{torn2004}, in this 
situation matter can not be further deposited and the accretion process 
comes to a halt. The thermal energy excess in the accreted layers 
produced by the previous very fast accretion begins to be removed via inner 
thermal diffusion. Hence, the WD contracts and recedes from the critical rotation 
condition on the thermal timescale of the hot and expanded surface layers. 
As a consequence, matter and angular momentum can be deposited once again and 
the accreting WD reacts by expanding and adopting an overcritical configuration. 
This implies that accretion occurs not as a continuous process but as isolated 
episodes and, hence, the resulting ``effective'' accretion rate onto the WD 
decreases. We remark that the exact average value of \mdote is determined by the 
conditions that the angular 
velocity at the interface star/disk is close to the critical keplerian value. 
Since the process is driven by the thermal response of the WD to mass and 
angular momentum deposition, PGIT defines this accretion regime as 
``self-regulated''. Two major consequences arise: 1) the off-center of C-burning
is avoided and 2) the total angular momentum of the WD continuously increases. 
As a matter of fact, the accreting object becomes a massive very fast rotator and it 
adopts a triaxal configuration (Jacobi ellipsoid). In this condition, 
gravitational waves can be emitted. The angular momentum loss stabilizes the 
accretion rate so that the WD increases its 
mass at an almost constant rate up and beyond the canonical \mch. Also in 
this case the star self-regulates the amount of matter which can be accreted, 
since the angular momentum losses depend on the ratio of rotational over 
the gravitational energy of the accreting WD \citep{fried1975}.

\cite{saio2004} generalized the work by PGIT, as differential rotation is 
accounted for, and according to 
their simulations off-center C-ignition occurs in any case. 
The difference in the final result is mainly determined by a 
different assumption on the accretion process. In particular, both the two 
groups find that, due to the deposition of matter and angular momentum, the 
accreting WD experience very soon the Roche instability. PGIT describe the following 
evolution by assuming that matter is accreted at an effective accretion rate 
smaller than the initial value and that the accreted matter carries angular 
momentum in any case. \cite{saio2004} assume that, when the 
surface rotational velocity is nearly critical, angular momentum can be 
transferred back from the WD to the disk via magnetic torque 
\citep{pacz1991,popha1992}, while 
\mdote maintains its initial high value. Since the rule 
of thumb is that high accretion rate drives to off-center C-ignition (valid also 
for rotating WDs) this is obtained very soon.

\cite{yoon2007} followed the merging process of a 0.9+0.6 \msune CO WDs initial 
binaries with a SPH simulation and use the resulting final configuration as input
for modeling the long thermal phase of mass accretion. Their approach is more 
refined than those in PGIT and \cite{saio2004} as they consider a central 
stellar-like object made by a compact, cold isothermal core rotating at slow 
rate surrounded by an hot extended envelope where angular velocity rapidly 
increase up to its breakup value. A centrifugal-supported
disk provides the reservoir for matter deposition onto the inner object. 
In their computation \cite{yoon2007} assume that, if the surface of the 
accreting core is rotating at a sub-keplerian angular velocity, angular momentum 
is deposited by the accreted matter, otherwise it is set to zero. Moreover, 
they allow the dissipation of angular momentum from the WD to the disk through 
all the accretion process. They finally obtain that off-center C-ignition 
is unavoidable for rapid accretion process, while a reduction of \mdote 
automatically drives to the onset of a thermonuclear runaway in the center once 
all the matter in the disk has been accreted onto the surviving stellar component. 
In this case rotation allows to obtain a spread in the total mass of the exploding 
object. As for the computations performed by \cite{saio2004}, the results 
by \cite{yoon2007} depend on the assumption that at the 
onset of the Roche instability angular momentum is no longer added but it is 
subtracted from the accretor. 

Our evolutionary models have been computed with an updated version of 
the 1D hydrostatic lagrangian evolutionary code FRANEC, firstly described 
in \cite{chie1989} and then revised in \cite{chie1998}.
The hydrostatic approximation adopted in the present work is suitable for 
our purposes, since we model the secular evolution of the accreting WD by 
assuming that matter flows from the newborn disk.

As in \cite{pie2003a} and \cite{pie2003b} we adopt the equation of state computed 
by \cite{stra1988} and successive updates \citep{prada2002}. 
We use the tables of radiative opacity provided OPAL group \citep{igles1996}  for
$T < 5\times 10^8$ K, while at higher temperature we adopt the ones derived from the 
Los Alamos Opacity Library \citep{huebner1977}. The contribution of the 
electron conduction to the total opacity is included according to the 
prescription by \cite{potek1999}.

The chemical composition of the surface 
layers of all the computed model is equal to $X_{^{12}C}$=0.506 
$X_{^{16}O}$=0.482 and $X_{^{22}Ne}$=0.011 while heavier elements 
are assumed to have solar abundances. This chemical pattern is 
representative of matter having experienced He-burning in a shell and 
is used as chemical composition of the accreted matter. 

As discussed in \cite{pie2003a}, the effects of rotation are included 
into the stellar structure equations by adopting the approximation 
by \cite{kipp1970} for the centrifugal potential, {\it i.e.} by 
averaging over a sphere the radial component of the centrifugal force. 
The effective gravity in the hydrostatic equilibrium equation is 
computed as the product of the local gravity and the factor $f$, defined 
as the ratio of mean centrifugal force and the gravitational one per unit 
mass. In formula:
\begin{equation}
{\frac{dP}{dr}}=-{\frac{GM}{r^2}}\rho\left(1-f\right)
\label{e:hydeq}
\end{equation}
As in \cite{pie2003a} we neglect the corrective factor in the expression 
for the radiative temperature gradient \citep[but\ see\ the\ discussion\ in][]{endal1976}.
As discussed in \cite{yoon2004}, the 1D approximation in the computation of the 
effective local gravity is accurate for angular velocities not exceeding 
$\sim 60$\% of the local critical value. For larger value, the adopted 
numerical procedure underestimates the centrifugal force. The consequence 
of such a limitation is discussed in \S \ref{s:evolu}. 

We assume that accreting WDs are rigid rotators, since they are fairly 
chemical homogeneous compact objects so that angular momentum 
redistribution could occur on a very short timescale as compared to 
the evolutionary timescale \citep{pie2003a,maeder2000}.
Moreover, as discussed in \cite{piro2008}, the occurrence of baroclinic instabilities 
and/or the shear growth determined by small magnetic fields determine 
a torque large enough to enforce rigid rotation in accreting WDs. 
Therefore, the time evolution of the angular velocity for accreting WDs 
is provided by the following equation:
\begin{eqnarray}
\omega(t)={\frac{J_0+J_{accr}}{I(t)}}
\end{eqnarray}
where $J_0$ is the initial WD angular momentum after the merging process, 
$J_{accr}$ is the amount of the angular momentum deposited by the 
accreted matter, and $I(t)$ is the momentum of inertia. 
For $J_{accr}$ we adopt the following expression
\begin{equation}
J_{accr}=\Delta M\cdot {\frac{2}{3}}R_{WD}^2\omega_{cr}
\end{equation}
where $\omega_{cr}$ is the critical value of the angular velocity at the 
surface, $R_{WD}$ the surface radius and $\Delta M$ the amount of matter 
accreted at each time step.
The determination of the angular velocity value is coupled to the solution of 
the equations for the stellar structure at each time step in order to take 
into account the feedback of the changes in the radius coordinate onto the 
$\omega$ value and, hence, on the factor $f$. 

If during the evolution the centrifugal force acting on a generic mass 
element of the star exceeds the local gravitational force 
(corresponding to the condition $f\ge 1$) the Roche instability occurs 
and the elements acquires a radial velocity larger than the escape 
velocity. Since $\omega_{cr}$ is a monotonically decreasing function of 
the radius and rigid rotation has been assumed, the Roche instability 
first occurs on the surface layer of the accreting WDs. 
Then, the amount of matter (and angular momentum) accreted at each time step 
is fixed by the condition that the star remain gravitationally bound ($f<1$ 
at the surface). 

The possibility that the equipotential surfaces could deform as the 
accreting WD spins up due to the continuous deposition of angular 
momentum can be checked by computing the quantity 
$\gamma=E_{rot}/E_{grav}$, where $E_{rot}$ and $E_{grav}$
are the total rotational and gravitational energy of the star, 
respectively. 
If such a quantity becomes larger than $\gamma_c\sim 0.1375$ \citep{fried1975} 
GWR are emitted carrying away angular momentum from 
the star at a rate given by 
\begin{eqnarray}
{\frac{dJ}{dt}}= -{\frac{J_*}{\tau_{GWR}}}e^{-t/\tau_{gwr}}
\label{e:loss}
\end{eqnarray}
where $J_*$ is the actual total angular momentum of the star and $\tau_{GWR}$, 
the characteristic timescale for the GWR emission, is computed according 
to \cite{fried1975} and \cite{chan1970}.  

When all the matter available in the system has been transferred to the WD or when 
the WD itself approaches the Chandrasekhar mass limit for rigidly rotating objects, 
we assume that the evolution is driven by the angular momentum loss from 
the star \citep{pie2003b}. In this case the efficiency of various physical 
mechanisms currently at work is described by a characteristic timescale $\tau$ 
and the corresponding amount of angular momentum subtracted at each time step 
is computed according to Eq. (\ref{e:loss}).
\begin{figure*}
 \includegraphics[width=15 cm]{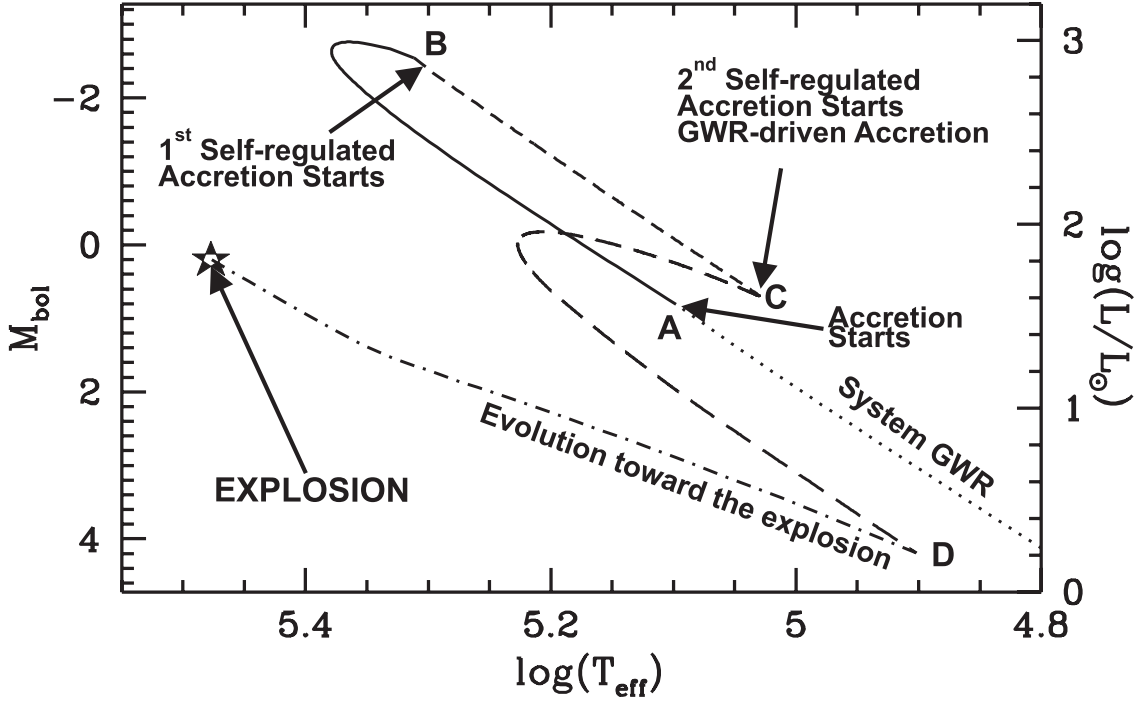}
  \caption{Evolution in the HR diagram of a CO WD with initial mass M=0.8 \msune 
   accreting CO-rich matter from a thick disk formed by the 0.7\msune 
   CO WD companion during the merging (see text for details).}
  \label{fig1}
\end{figure*}

\section{Evolution Toward the Explosion}\label{s:evolu}

In the following we will refer to the evolutionary outcomes of the self-regulated 
accretion scenario for merged DD systems originally presented in PGIT. 
Following their prescriptions for the evolution of the merged object, 
we compute a number of additional models 
to better describe the broad spectrum of physical situations expected to occur in 
nature. 
In the present work we focus only on the behaviour of the accreting WD during 
the evolution to the explosion, not accounting for the contribution of the 
accretion disk to the observational features.

The evolution in the HR diagram along the accretion phase of a CO WD with initial 
mass ${\mathrm M_{WD}=0.8}$\msune is reported in Fig. \ref{fig1}. The CO rich 
material flows directly from the disk formed by the disruption of a 0.7\msune CO companion 
during the merging phase. 

We recall that during the pre-merging phase the two components interact tidally 
so that part of the rotational energy of the system is employed 
to maintain the synchronization of the orbits and to heat both the WDs. 
As a consequence, at the merging the two compact objects rotate very rapidly and, in addition, their 
temperature profiles are largely modified relative to those of the initial structures. 
This implies that just before the merging the two WDs evolve in the HR diagram along the 
cooling sequence toward larger surface luminosities and temperatures 
(dotted line in Fig. \ref{fig1} up to point {\sl A} for the more massive component).
The evolution towards the merging is driven by the emission of gravitational 
wave radiation, which extracts angular momentum from the system, determining 
the spiral-in of the two components \citep{iben84} and giving rise to first
detectable GWR signature \citep{nel2001,nel2004,yu2010}.
Such an occurrence is labeled in Fig. 1 as {\sl System GWR}.

The post-merging system is composed by the initially more massive component and by a 
disk formed by the disruption of the initially less massive CO WD. 
\cite{lig2009} analyzed in detail the merging process by means of hydrodynamical 
computations. They find that, if the mass ratio of the WDs is not close to 1, the 
newborn disk is rather thin and very extended. Moreover, as already mentioned, 
off-center C-burning ignited 
at the WD-disk interface as soon as the merging starts, is rapidly quenched.
Under these conditions, matter is expected to flow from the disk to the surviving 
CO WD depositing both angular momentum and energy. 
\cite{lig2009} are not able to determine the effective accretion rate from
the disk to the WD, even if they tend to prefer a high value of \mdote because
of the instabilities affecting the disk. In such a case, if no other physical mechanism 
is at work, accreted carbon would be ignited very soon in the accreted layers and 
C-burning should steadily propagate toward the center, producing an O-Ne-Mg WD, which 
eventually will collapse to a neutron star, avoiding a thermonuclear explosion.
On the other hand, according to the DD rotating scenario the effective \mdote 
is regulated by the thermal and angular momentum content of the accreting WD. 
In particular, accretion occurs when the WD is in a condition to accept matter and it lasts for 
the time the physical properties of the accretor allows it to occur, as discussed in the 
following \citep[see\ also\ the\ discussion\ in][]{jor2012}.

We distinguish four different phases in the evolution of the merged object:
\begin{enumerate} 
\item {\bf constant $\mathrm{\dot{M}}$ phase}: initially matter flows from 
the disk to the WD at a very high rate ($\mathrm{\dot{M}\ge 10^{-5}}$\msunyrm). 
Due to the large amount of thermal energy delivered by the accreted matter, 
the external layers of the accreting WD heat up, while the surface radius 
greatly increases as a result of both the local thermal 
excess and the rapid spinning up. Hence, both the surface luminosity and temperature 
increase (solid line in Fig. \ref{fig1}). When expansion becomes the leading 
parameter of the evolution, the external layers begin to cool and the surface 
luminosity decreases;
\item {\bf first self-regulated accretion phase}: as the accreting WD spins up, due to 
the deposition of angular momentum by the accreted matter, and the external 
layers expand, the critical rotational velocity is attained for the first time 
(point B in Fig. \ref{fig1}). 
As a consequence, matter can no longer be accreted and mass transfer comes to a 
halt for a while. Thermal energy continues to flow from the surface layers toward 
the inner zones. The consequence of this new situation is that the WD contracts 
and recedes from the critical condition, 
so that accretion can resume again and proceed up to when the critical 
rotational velocity is attained once more. 
During this phase the accretion process is therefore driven by the thermal 
diffusion timescale of the accreting object. As a matter of fact, 
the rate at which matter is effectively deposited onto the WD reduces with time and, hence, 
also the surface luminosity decreases (dashed line in Fig. \ref{fig1}). In fact,
the thermal transfer time scale becomes longer since the thermal gradient
between the surface and the inner layers is reduced with time.
This means that it becomes harder and harder for the star to recede from the 
critical condition.
\item {\bf GWR-driven accretion phase}: should no other phenomenon occur, accretion
would come to a definitive halt. But the continuous deposition of angular momentum 
by the accreted matter, increases the rotational energy until the latter 
exceeds a critical fraction of the gravitational energy (point C in Fig. 
\ref{fig1}). In this condition gravitational wave 
emission sets in and lasts for a long time. GWR allows the accreting 
WD to constantly recede from the critical condition, thus accreting matter (and
momentum) at an 
almost constant accretion rate. This corresponds to a continuous decrease of the 
surface radius and to a corresponding heating of the surface layers via compression. 
Then, the surface luminosity increases once again 
(long-dashed line in Fig. \ref{fig1}). Later on, when a large part of the thermal 
energy locally stored in the accreted layers is removed via inward thermal 
diffusion the surface luminosity decreases. When the accreting WD approaches
\mche for rigidly rotating degenerate objects, the secular stability is attained 
once again and GWR emission ceases (point D in Fig. \ref{fig1}). Hence the accretion 
rate rapidly decreases and matter cannot be any longer accreted. 
{\bf During all this phase the accreting WD may deviate slightly from 
axisymmetry around the principal rotation axis. This implies that the WD acquires 
both a poloidal and an equatorial ellipticity, thus producing two ``spectral line''
in the emitted gravitational radiation. We consider only the gravitational 
emission related to the ellipticity in the equatorial plane, whose frequency is equal to 
the double of the spinning frequency of the WD \citep{zim79}. The amplitude of this 
signal is computed according to \cite{sha+teu1983} as:
\begin{equation}
h_0=\frac{G^2}{c^4} \frac{M_{WD}^{2}}{R_{WD}}\frac{1}{d}
\end{equation}
where $d$ is the distance from the source and $M_{WD}$ and $R_{WD}$ the actual value of
mass and radius of the accreting object, respectively. 
}
The time evolution of the GWR characteristics are reported in Fig. \ref{fig2} 
where we plot also the time evolution of the surface luminosity. 
\begin{figure}
 \includegraphics[width=\columnwidth]{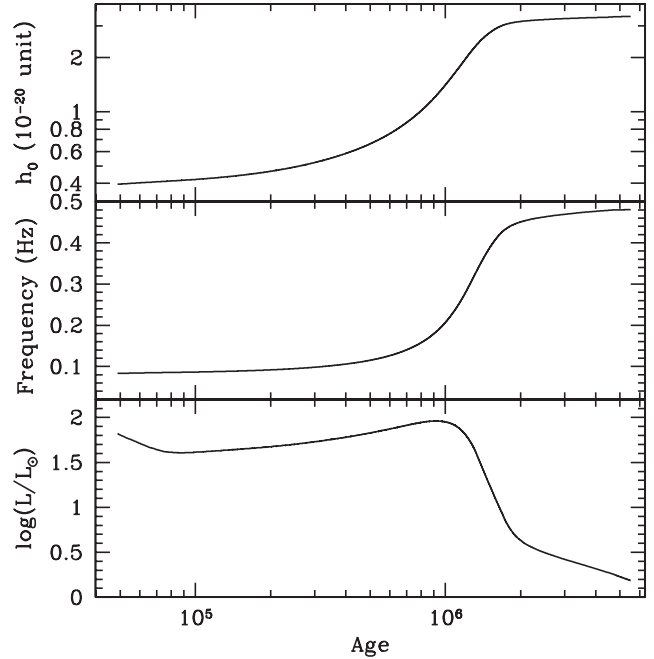}
  \caption{Some physical properties of the same model as in Fig. \ref{fig1}. 
   {\sl Upper panel}: time evolution of the GWR amplitude at a distance of d=1 kpc 
   from the source. {\sl Middle panel}: time evolution of the GWR frequency. 
   {\sl Lower panel}: time evolution of the surface luminosity of the accreting WD.} 
  \label{fig2}
\end{figure}
\item {\bf Evolution toward the explosion}: the final part of the evolution is triggered 
by the loss of angular momentum from the accreted WD; this implies 
that the structure contracts homologously and heats up by compression up to the explosion.
As a consequence, the surface luminosity increases (dot-dashed line in Fig. \ref{fig1}).
During this phase, the WD reacts to the angular momentum loss by spinning
up. 
This is clearly shown in Fig.\ref{fig3}, where we report the evolution of the 
angular velocity as a function of time after the accretion halts and up to the 
C-ignition at the center. For the model with $\tau=10^5$ yr the rotational velocity 
increases from 1.51 rad s$^{-1}$ (point D in Fig. \ref{fig1}) to 3.19 rad s$^{-1}$
(point ``Explosion'' in the same figure).
Such an occurrence can be easily explained by observing that the contraction triggered by the 
angular momentum loss produces the reduction of the momentum of inertia of the star
and hence the increase of angular velocity due to the conservation of total angular 
momentum \citep{ger00,bos12}. 
We note that \cite{ilkov2012} analyze in detail this evolutionary phase up to the 
explosion for a differentially rotating WD formed in the framework of the Core 
Degenerate scenario \citep{kashi2011}. They conclude that the emission of 
gravitational wave radiation in the $r$ mode is inefficient at all to drive the 
merger remnant to the explosion and propose as effective braking mechanism the 
emission of magneto-dipole radiation. 
{\bf In our models we fix the value of $\tau$ according to the analysis performed 
by PGIT, by assuming an intermediate efficiency of the physical processes 
driving the angular momentum losses. The angular momentum is subtracted from the 
WD according to Eq. (\ref{e:loss}).}
\end{enumerate}
\begin{figure}
 \includegraphics[width=\columnwidth]{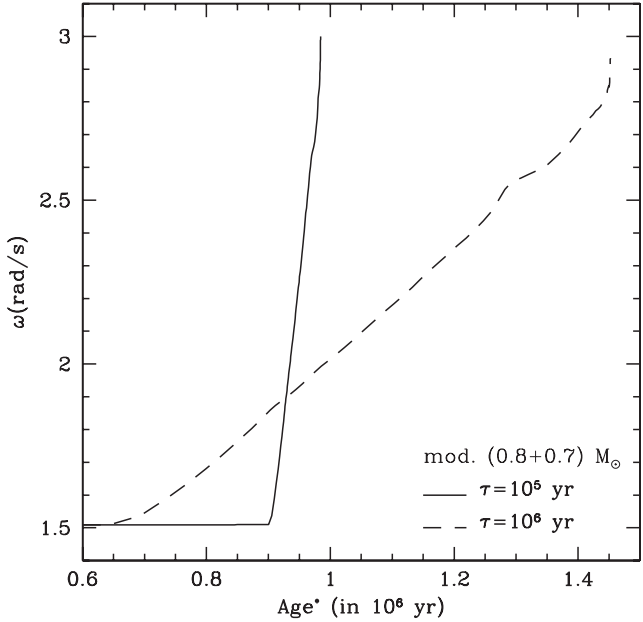}
  \caption{Time evolution of the angular velocity of the accreted WD during the 
  last part of the evolution driving to the explosion for the model (0.8+0.7)\msun . 
  The abscissa ($Age^*$) represents the time elapsed from the halt of the 
  accretion process up to ignition of C-burning at the center. The two lines refer 
  to model computed with different braking efficiencies, as labeled inside the 
  figure (for more details see \S \ref{s:evolu} and \S \ref{s:synthe}).} 
  \label{fig3}
\end{figure}

\begin{figure}
\includegraphics[width=\columnwidth]{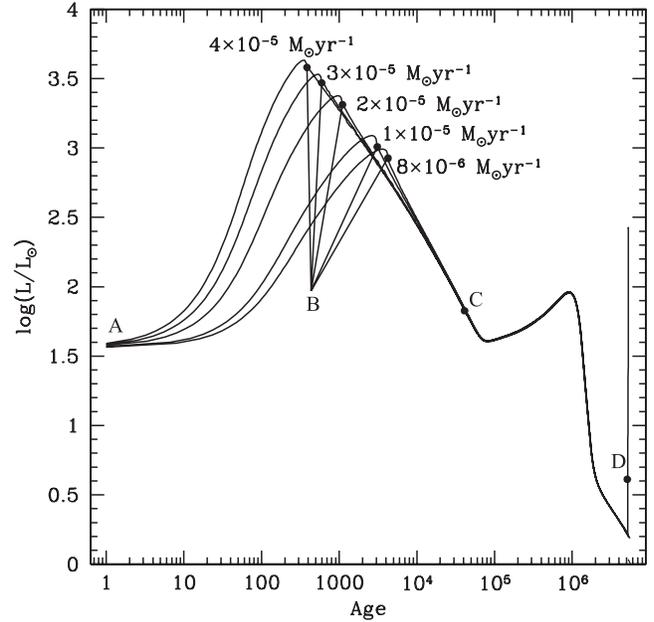}
\caption{Time evolution of the surface luminosity of a CO WD with initial mass 
0.8 \msune and different values of \mdote (as labeled) at the beginning of mass 
accretion process from the disk to the WD. The filled dots 
mark the onset of the various accretion regimes: 
A - constat \mdot; 
B - self-regulated accretion phase; 
C - GWR-driven accretion phase; 
D - Final path to explosion. 
For more details see text.} 
\label{fig4}
\end{figure}

We note that the rotating DD scenario is 
the direct consequence of the self-consistent inclusion of rotational effects 
in the accretion process. The only free parameter in the model is 
the initial value of the accretion rate after the merging process, which affects
just the duration of the constant \mdote and self-regulated accretion
phases and the maximum attained luminosity, leaving unaltered the final result. 
This is clearly shown in Fig. \ref{fig4}, where we report the time evolution of the 
surface luminosity for models with the same initial WD mass and thermal structure 
as in PGIT, but with different values of \mdote for 
the post-merged configuration, as labeled inside the figure. 
In Tab. \ref{tab1} we report for each model in Fig. \ref{fig4} the time duration 
of the four different evolutionary phases up to the explosion. 

It is important to note also that the main features of the PGIT model 
({\it i.e.} the self-regulation of the effective accretion rate and the 
emission of GWR) remain unaltered also when taking into account both the 
limitation of computing 1D models and the possible uncertainty in the amount of 
angular momentum effectively deposited after the onset of the critical rotation 
condition ($\omega_{WD}\simeq \omega_{cr}$). 
In particular, when considering that in 1D models the local effective gravity is 
overestimated in stellar layers with $f > 0.6$, it turns out that the corresponding 
surface radius of the accretor is underestimated. This implies that the WD attains 
sooner the critical condition, but such an occurrence does not affect the following 
evolution, which is driven by the thermal response of the accretor to mass and 
angular momentum deposition. 
\begin{table}
 \caption{Time duration of the various accretion phases experienced by a 0.8 
          \msune CO WD accreting CO rich matter as a consequence of a merging 
          with its degenerate companion of 0.7 \msune as a function of 
          different initial \mdot.}
 \label{tab1}
 \begin{tabular}{lrrrr}
   \hline
    \mdot & ${\mathrm \Delta T_{AB}}$ & ${\mathrm \Delta T_{BC}}$ & 
                      ${\mathrm \Delta T_{CD}}$ & ${\mathrm \Delta T_{DE}}$ \\
    (\msunyrm)& (yr) & (yr)& (yr) & (yr) \\
   \hline
    $8\times 10^{-6}$ & 4281 & 38524 & 5119599 & 84257 \\
    $1\times 10^{-5}$ & 3096 & 40171 & 5205641 & 83863 \\
    $2\times 10^{-5}$ & 1104 & 41422 & 5212430 & 83308 \\
    $3\times 10^{-5}$ &  595 & 41700 & 5079911 & 89886 \\
    $4\times 10^{-5}$ &  385 & 41658 & 5457113 & 89856 \\
  \hline
 \end{tabular}

 \medskip
 {\em AB}: Evolutionary phase with \mdot=cost.; 
 {\em BC}: Self-regulated accretion phase; 
 {\em CD}: GWR-driven accretion phase; 
 {\em DE}: Evolution up to the explosion.
\end{table} 

Similar considerations can be done also when assuming that a smaller amount of 
angular momentum is deposited by the accreted matter after the onset of the 
critical rotation condition. In fact, the angular momentum increases at a lower 
rate and the condition $\gamma\ge\gamma_c$ is attained later, so that the 
self-regulated accretion phase lasts for a longer time. 
However, when gravitational waves start to be emitted, the further evolution is 
driven by the GWR timescale and, hence, the average value of \mdote remains 
practically unaltered as well as the time up to end of the accretion process. 

This scenario well describes also the evolution of differentially rotating CO WDs which 
accrete matter from a CO rich disk, as conjectured in \cite{pal2009}. In fact, the 
time evolution of the accretion rate is exactly the same, since it is determined by 
the same physical processes, while the duration of the GWR-driven accretion phase can 
increase by a factor up to 2-3, depending on the total mass of the system 
at the merging time \citep[from\ 1.4\ up\ to\ 2.4\ \msune as\ discussed\ in][]{pal2009}. 
On the other hand, in differentially rotating structures, after the accretion comes to 
a halt due to the lack of matter in the disk, the evolution toward the explosion is 
triggered by the inner angular momentum redistribution via viscous shear, occurring 
on a timescale of the order of $\mathrm{10^{5}\div 10^{6}}$ yr. 

It is important to note that the evolutionary path previously described 
is followed also by DD systems with total mass smaller than the standard non-rotating 
Chandrasekhar limit; in this case, when all the matter in the disk has 
been transferred to the surviving component, the following evolution 
does not lead to an explosion but to a cooling massive WD. 
In order to make more clear this point we present additional evolutionary 
sequences of merged DD systems with different total mass and mass of the primary 
component, namely DD systems made by a (0.6+0.5) \msun , (0.7+0.6) \msun , (0.9+0.6) 
\msune and (1.0+0.5) \msune CO WDs. 
\begin{figure}
\includegraphics[width=\columnwidth]{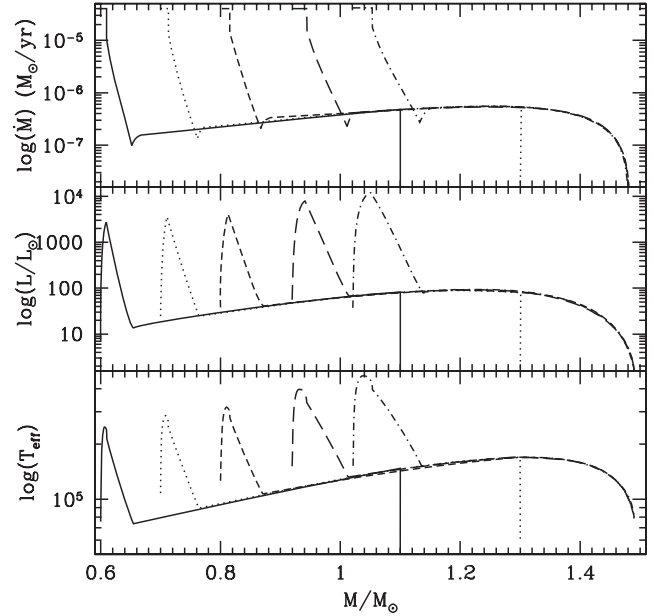}
\caption{Evolution, as a function of the total mass of the accreting WDs, of the effective 
   accretion rate ({\sl upper panel}), of the surface luminosity ({\sl middle panel}) 
   and of the effective temperature ({\sl lower panel}). The initial accretion rate 
   for the post-merging objects has been fixed to \mdot$=4\times 10^{-5}$\msunyrm . 
   Solid lines refer to the (0.6+0.5) \msun , dotted lines to the (0.7+0.6) \msun , 
   dashed lines to (0.8+0.7) \msun , long-dashed lines to the (0.9+0.6) \msune and 
   dot-dashed lines to the (1.0+0.5) \msune models, respectively.}
\label{fig5}
\end{figure}

In our computations we follow the same prescriptions as in PGIT, but we adopt for the 
initial phase with constant \mdote the value of $\mathrm{4\times 10^{-5}}$ \msunyrm. 
The results are summarized in Fig. \ref{fig5}, where we report 
as a function of the accreting WD mass the effective accretion rate (upper panel), the 
surface luminosity (middle panel) and the effective temperature (lower panel). The 
sudden drop of the curves in the three panels at M=1.1 and M=1.3 \msune marks the end 
of the accretion process due to the exhaustion of the matter reservoir in the 
accretion disk for models (0.6+0.5) \msune (solid line) and (0.7+0.6) \msune 
(dotted line), respectively. Fig. \ref{fig5} also discloses that the 
physical properties during the GWR-driven accretion phase are largely 
independent of the initial mass of the accreting WD. Moreover, 
since during this phase accreting WDs are rotating with angular velocity equal to 
the critical one, which depends only on the surface radius, it transpares that 
the amplitude and the frequency of the emitted gravitational radiation 
do not depend on the initial value of $\mathrm{M_{WD}}$, but depend only on the actual 
value of the total mass. 
\begin{table}
 \caption{Time duration of the various accretion regimes experienced by models with 
 different total mass and mass of the accretor, by fixing the values of \mdote during 
 the constant accretion rate phase at $4\times 10^{-5}$\msunyrm. 
 For the sake of completeness, we report also the data relative to the model 
 (0.8+0.7) \msun .}
 \label{tab2}
 \begin{tabular}{lrrrr}
   \hline
    Model & ${\mathrm \Delta T_{AB}}$ & ${\mathrm \Delta T_{BC}}$ & 
                      ${\mathrm \Delta T_{CD}}$ & ${\mathrm \Delta T_{DE}}$ \\
             & (yr) & (yr)& (yr) & (yr) \\
   \hline
    (0.6+0.5) & 245 & 58960 & 1815793 &    -    \\
    (0.7+0.6) & 318 & 52203 & 1513552 &    -    \\
    (0.8+0.7) & 385 & 41658 & 5457113 & 89956 \\
    (0.9+0.6) & 607 & 52105 & 5099653 & 84190 \\
    (1.0+0.5) & 802 & 55865 & 4840686 & 84190 \\
  \hline
 \end{tabular}
\end{table} 

In Tab. \ref{tab2} we report the time duration of the accretion regimes experienced 
by these newly computed models. Since the (0.6+0.5) and (0.7+0.6) \msune models do not 
attain the physical conditions suitable to produce a SN Ia event as the final total 
mass of the accreting WDs is smaller than the standard non-rotating \mch, 
$\Delta T_{DE}$ can not be defined. 

\section{Expected Frequency of Pre-Exploding Systems}\label{s:synthe}

In the following we estimate the expected number of pre-explosive systems in the 
Galaxy for each evolutionary phase after the formation of the accretion disk 
around the originally more massive CO WD and up to the explosion.
We adopt the duration of various phases leading to the explosion of models 
initially accreting at \mdot=${\mathrm 4\times 10^{-5}}$\msunyrm \ as listed in 
Tab. \ref{tab2}. 
By an inspection of Tab. \ref{tab1}, it comes out that such a choice for 
the initial value of the accretion rate does not imply appreciable differences 
in the evolutionary timescales, except for the constant \mdote phase, which,
in addition, is the shortest one.
For the final approach to explosion we fix the braking timescale to 
${\mathrm \tau_{B}=10^{5}}$ yr. Such an assumption may affect to some extent the 
estimated frequency, as discussed below in more detail.
Finally, following \cite{bra2011}, we assume that the simmering phase, lasting 
from the crossing of the ignition curve up to the crossing of the dynamical curve, 
is equal to ${\mathrm \Delta T_{simm}\simeq 530}$ yr\footnote{The evolutionary time 
elapsing from the onset of the thermonuclear runaway up to the moment when 
the nuclear timescale becomes comparable to the WD sound crossing time 
is usually addressed as ``simmering''. During this phase, the combined action 
of convective mixing, carbon burning and e-captures largely modifies the chemical
composition of the WD, thus determining to some extent the observational properties 
of the corresponding Supernova event.}.
Since we focus our attention on the Galaxy, we assume an expected SN Ia rate 
of one event every 200 yrs \citep[see,\ {\it e.g.},][]{cap1999,li2011a}.

In Tab. \ref{tab3} we report the expected number of merged DD systems good candidates 
for SNe Ia progenitors for each evolutionary phase up to the explosion, 
the total being $\sim$ 28000.
As already mentioned in the previous section, merged systems with initial total 
mass smaller than \mche experience the same evolution. Hence, they populate the 
initial parts of the same 
evolutionary phases, even if they end their life as massive CO WDs. On general 
grounds, it can be estimated that the number of merged DD systems less massive 
than \mche is only slightly larger (just $\sim$ 1.5) than those with larger mass
\citep{IT1985,tor89}. The properties of this 
``SN manque'' component deserve a more complex analysis via detailed population
synthesis computations, which is far beyond the aim of the present work. So in the 
following we consider only those merged DD systems with total mass 
larger than \mch; in this way we properly derive the properties of the 
binary population we are considering, even if we underestimate the number of 
objects contributing to each phase. 
{\bf In our computation we assume that all the merged DD systems are uniformly distributed 
in the disk of the Galaxy.}
\begin{table}
 \caption{Expected number of merged DD systems which will produce a SN Ia event 
          in the Galaxy along various pre-explosion phases, as reported 
          in the first column.}
 \label{tab3}
 \begin{tabular}{lr}
   \hline
    Evolutionary Phase & Merged DD systems\\
   \hline
    constant \mdot           &        2  \\
    Self-regulated accretion &      208  \\
    GWR-driven accretion     &    27286  \\
    final path to explosion  &      421  \\
    Simmering                &        3  \\
  \hline
 \end{tabular}
\end{table} 

\begin{figure}
\includegraphics[width=\columnwidth]{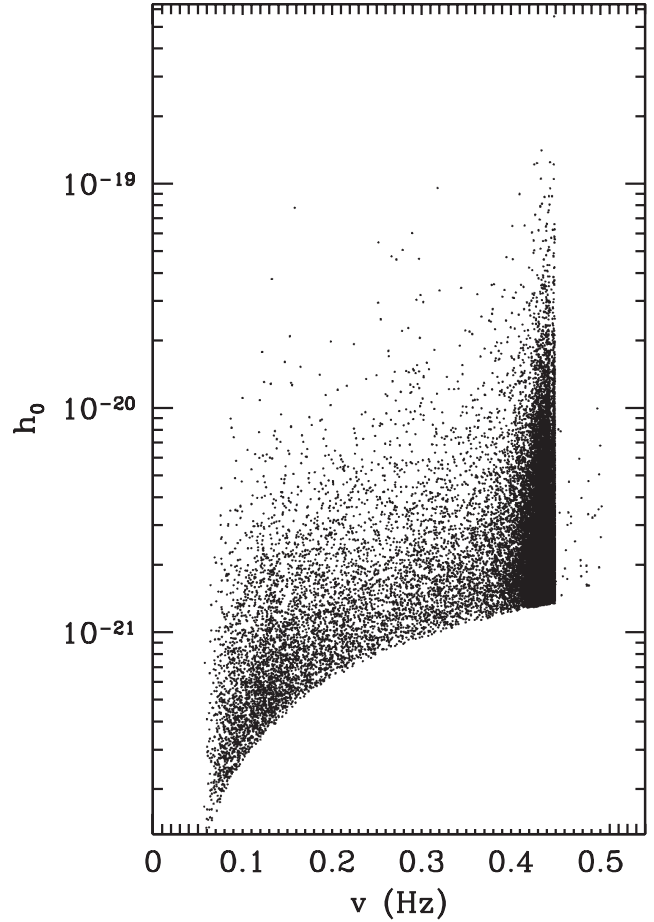}
\caption{Distribution in the $\nu - h_0$ plane of all DD merged systems good 
candidates as SNe Ia progenitors.} 
\label{fig6}
\end{figure}

In Fig. \ref{fig6} we report in the $\nu - h_{0}$ plane all the objects (27286) 
currently in the GWR-driven accretion phase. The sharp border at $\nu\sim 0.06$ 
Hz corresponds to systems with lower mass, close to the standard non rotating 
\mch, and which have just entered in the GWR-driven accretion regimes, 
while the other one at $\nu\sim 0.45$ Hz corresponds to systems which are attaining 
the limiting mass for the rigidly rotating degenerate objects, 
the very few objects ($\sim$ 50) with $\nu > 0.45$ Hz having already attained 
such a limit. 
The lower limit in $h_0$ represents the Galactic accreting systems furthest 
from the Earth. 
The clump of systems at high frequency ($\nu >$ 0.4 Hz) can be easily explained 
by noting that the last part of the GWR-driven accretion phase for massive 
DD systems is very long (see Fig. \ref{fig2}) since the rate at which 
matter is effectively accreted onto the WD rapidly decreases (see Fig. \ref{fig5}). 

\begin{figure}
\includegraphics[width=\columnwidth]{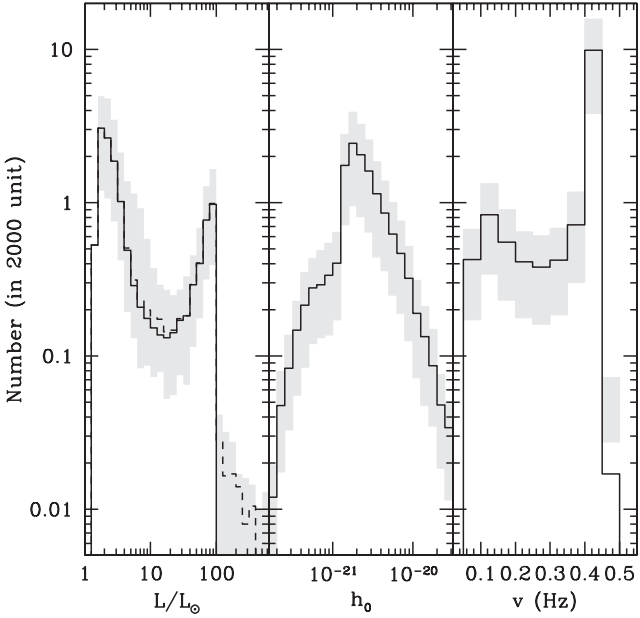}
\caption{From left to right we plot the expected number of merged DD systems 
with $M_{\mathrm tot}>$\mche (normalized to 2000) 
experiencing the GWR-driven accretion phase as a function of the surface 
luminosity, amplitude of the emitted gravitational signal and corresponding 
frequency, respectively. The dashed line in the left panel refers to all the post 
merged system. The shaded region represents the uncertainty in the estimated numbers 
(for more details see text).
} 
\label{fig7}
\end{figure}

In Fig. \ref{fig7} we report the expected number of objects experiencing 
the GWR-driven accretion phase as a function of the surface luminosity (left panel, solid line), 
of the GWR amplitude (middle panel) and of the corresponding frequency (right panel). 
As it can be noticed, only $\sim$25\% of these systems have a high luminosity 
(let say larger than $\mathrm {log(L/L_\odot)=1}$), while louder GWR emitters 
($h_0\ge 3.2\times 10^{-21}$) with larger frequency ($\nu\ge 0.44$ Hz) corresponds 
to $\sim$ 30\% of the total number. It is important to remark that louder GWR 
emitters correspond to less luminous objects.

\begin{figure}
\includegraphics[width=\columnwidth]{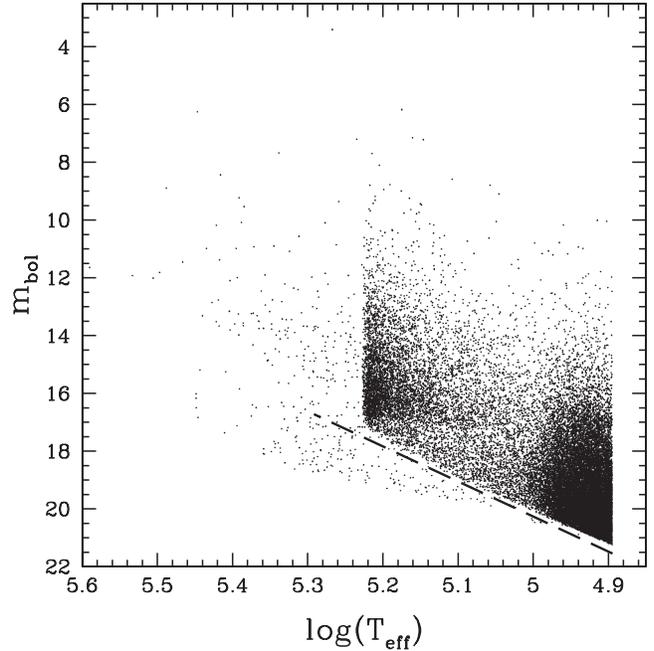}
\caption{Distribution in the $T_{eff} - m_{bol}$ plane of all the Galactic
post-merged DD systems with total mass larger than \mch. For more details see text.} 
\label{fig8}
\end{figure}

In Fig. \ref{fig8} we plot the expected distribution in the apparent bolometric 
magnitude (${\mathrm m_{bol}}$)-effective temperature plane of all the 
post-merged systems, good candidates for SNe Ia events. 
Systems currently experiencing the GWR-driven accretion phase have 
$T_{eff}\le$ 178000 K and apparent bolometric magnitude above the 
long-dashed line. Systems above this lines and with 
$126000\le T_{eff}\le 400000$ K are experiencing the self-regulated accretion phase, 
while objects not currently accreting but evolving to the explosion are almost 
uniformly distributed in the bolometric magnitude range 14-20 and effective temperature
range 90000 - 250000 K. 
The contribution of the latter component to the luminosity 
function of objects currently evolving toward an explosion of SN proportion 
is illustrated by the dashed line in the left panel of Fig. \ref{fig7}.

The main sources of uncertainty in the expected frequencies of merged DDs which will 
evolve into a SN Ia event are represented by the uncertainty in the SNe Ia rate of our 
Galaxy, which 
affects uniformly the number of systems in each evolutionary phase, and by 
the duration of the braking phase, which determines only the number of objects 
currently decaying to the ignition conditions. The former can vary in the range 
0.2 - 1 Supernova 
per century, as derived by taking into account the uncertainties both in the observed 
rate and in the luminosity in the B band of the Galaxy. The latter mainly depends 
on the efficiency of angular momentum redistribution along the accreted WD and/or 
angular momentum loss by the star itself. In such a case, according to the analysis in
PGIT, it can be derived that these processes occur on a time scale in the range 
$10^{5} - 10^{6}$ yr. 
By combining together these uncertainties we derive that the number of merged DD systems 
good candidate as SNe Ia progenitors currently in the Galaxy may range from 
$\sim$12000 to $\sim$45000, 
while the number of accreting objects emitting GWR varies in the range 11000 - 44000. 
The shaded region in each panel of Fig. \ref{fig7} represents the maximum uncertainties 
for the estimated number of Galactic DD systems, which have already experienced 
a merging and with total mass larger than \mch. 

\section{Discussion and Conclusions}\label{s:conclu}

The overall analysis of the evolutionary path leading to a thermonuclear explosion 
clearly discloses that both the Single Degenerate and Double Degenerate models for the 
progenitors of SNe Ia have severe problems. Up to now, the only possible solution to this 
impasse, at least for the DD scenario, is rotation, which represents, as first 
addressed by PGIT, the leading parameter regulating the accretion process. When 
rotation is accounted for, some important implications directly follow. 
First of all, the accreting WD can largely exceed the Chandrasekhar mass limit, the 
final value depending on the initial mass of the original DD system. This determines 
very easily a spread in the mass of the exploding objects, in the range 1.4 - 2.4 \msun. 
{\bf If it is assumed that more massive progenitors produce larger amount of 
${^{56}Ni}$ during the explosion, rotation could explain the 
differences in the observed magnitude at the maximum epoch in light curves of 
``normal SNe Ia'' as well as the rare, very bright events, like SN 2003fg 
\citep{how06}, SN 2007if \citep{sca10} and SN 2009dc \citep{sil11,tau11}.}
The second important consequence is that the explosion does not occur soon after 
all the matter in the disk has been accreted onto the WD, as rotation 
keeps the structure over-stable. The thermonuclear runaway can occur only when, 
owing to the redistribution or loss of angular momentum, a strong compression of 
the whole structure occurs, triggering the explosive C-burning at the center.
This corresponds to a radical change in our way of looking at thermonuclear Supernovae. 
In fact, the accreting WD does not attain a critical mass, but the value of the 
critical mass decreases up to the actual mass of the WD, thus determining the 
explosion. 
As a final consequence, it follows that it is practically impossible to find 
any trace of the progenitors of nearby recent SNe Ia in archival frames, as the 
explosion occurs several million years after the accretion has halted.
The previous considerations directly emerge by including rotation self-consistently 
in the modeling of SNe Ia progenitors and are valid for any kind of accreting scenario 
\citep[{\it e.g.}\ for\ the\ SD\ scenario\ see][]{hac12}. 

For this reason, according to the prescriptions in PGIT, we computed the evolution 
of merged DD systems by including the effect of rotation as the regulator of the 
accretion process.
Several models with different initial parameters have been evolved from the 
heating regime, just before the merging of the two components, up to few seconds 
before the explosion, when a large convective core has already developed in the 
accreting WD and the hydrostatic equilibrium assumption is no longer valid.

We have demonstrated that the only free parameter in rotating DD models, the initial 
value of \mdot , does not affect the following evolution, regulated by 
well established physical properties and free of any further assumption.
We showed that the accreting WD experiences three different accretion regimes, 
and we explained in detail their corresponding evolutionary properties.
We also analyzed the gravitational signal emitted during the third 
accretion regime. 

According to our results, the evolutionary properties of accreting WDs 
(luminosity, effective temperature, gravitational signal) depend only on the 
actual value of the total mass, but not on the initial parameters of the system
or on the previous accretion history. 

Using the computed models, we provide for the Galaxy an estimate of the expected 
number of merged DDs good candidates for SNe Ia progenitors currently emitting 
GWR and the properties of the gravitational signal in the $\nu - h_0$ 
parameter space as it should be observable from the Earth. We also provide the surface 
properties (apparent bolometric magnitude and effective temperature) of the 
accreting WDs. 

The gravitational signal produced by these systems is largely below the sensitivity curve 
of the current generation of space born interferometers. For example, the LISA 
apparatus will be able to detect with a S/N=1 only very few events (2-3), 
corresponding to objects with total mass larger than the canonical \mche ({\it i.e.} 
with $\nu\ge 0.4$) and located inside 60 pc from the Earth. However, this could represent 
an unambiguous confirmation of the pivotal role played by rotation in the evolution of 
merged DD systems, as well as a clear indication that a part of the type Ia events 
arises from DD systems. 

{\bf 
On the other hand, the next generation of space born interferometers, such as eLISA 
\cite[evolved LISA: Laser Interferometer Space Antenna,][]{pau12a,pau12b}
and
DECIGO \cite[Deci-Hertz Interferometer Gravitational-wave Observer,][]{decigo1},
will be able to detect the whole range of frequencies and amplitudes shown 
in Fig. \ref{fig6}, thus providing a deeper insight into SNe Ia. 
}
As a matter of fact, it will be possible to get information on the
actual spectrum of accreting WDs masses, thus shedding light on the efficiency of 
angular momentum transport in degenerate stars (differential {\it vs} rigid rotation). 
Moreover, the exact number counts of events as a function of the observed frequency 
and signal amplitude will allow one to determine if DD systems are the only progenitors 
of SNe Ia or if some additional contribution comes from different progenitor systems. 

\section*{Acknowledgments}
{\bf 
This paper has been inspected by two referees. Even if the second
referee was skeptic about our reference model and its basic assumptions, 
she/he provided us helpful comments to improve the manuscript and 
suggested additional observational tests to verify the reliability of our work.
We thank B. K\"ulebi and K.Y. Ek{\c s}i for insightful and stimulating 
discussions. 
We thank J. Danziger for carefully reading the manuscript and for comments and
suggestions. }

\bsp

\label{lastpage}

\end{document}